\documentclass[aoas,preprint]{imsart}

\usepackage{amsthm,amsmath,amsfonts}
\RequirePackage[colorlinks,citecolor=blue,urlcolor=blue]{hyperref}
\usepackage[round]{natbib}
\usepackage[english]{babel}
\usepackage{latexsym}
\usepackage{subfigure}
\usepackage{epsfig}
\usepackage{color,soul}
\usepackage{moreverb}
\usepackage{mathtools}
\usepackage{tikz, color}
\usepackage{enumerate,linegoal}
\usepackage{calc}
\usepackage{latexsym}
\usepackage{amsmath}
\usepackage{amsfonts}
\usepackage{amssymb}
\usepackage{bm}
\usepackage{psfrag}
\usepackage{graphicx}
\usepackage{epstopdf}
\usepackage{framed}
\usepackage{booktabs}
 \usepackage[flushleft]{threeparttable}

\usepackage{xfrac,array,booktabs}
\usepackage{verbatim}
\usepackage[shortlabels]{enumitem}
\usepackage{pdflscape}
\usepackage{rotating}

\startlocaldefs
\newcommand{\tp}{^{\top}}
\newcommand{\rd}{\mathrm{d}}
\newcommand{\rds}{\,\rd}
\newcommand{\expect}{\operatorname{E}}
\newcommand{\var}{\operatorname{var}}

\newtheorem{theorem}{Theorem}
\newtheorem{lemma}{Lemma}
\endlocaldefs

\begin{document}

\begin{frontmatter}

\title{Directional quantile classifiers}
\runtitle{Directional quantile classifiers}

\begin{aug}
\author{\fnms{Alessio} \snm{Farcomeni}\thanksref{t1}\ead[label=e1]{alessio.farcomeni@uniroma2.it}}
\and
\author{\fnms{Marco} \snm{Geraci}\corref{}\thanksref{t2}\ead[label=e2]{geraci@mailbox.sc.edu}}
\and
\author{\fnms{Cinzia} \snm{Viroli}\thanksref{t3}\ead[label=e3]{cinzia.viroli@unibo.it}}

\thankstext{t1}{Corresponding author: Alessio Farcomeni, Department of Economics and Finance, University of Rome ``Tor Vergata'', Italy. \printead{e1}}

\runauthor{Farcomeni et al}

\affiliation{University of Rome ``Tor Vergata''\thanksmark{t1}, University of South Carolina\thanksmark{t2}, University of Bologna\thanksmark{t3}}

\end{aug}

\begin{abstract}
\quad We introduce classifiers based on directional quantiles. We derive theoretical results for selecting optimal quantile levels given a direction, and, conversely, an optimal direction given a quantile level. We also show that the misclassification rate is infinitesimal if population distributions differ by at most a location shift and if the number of directions is allowed to diverge at the same rate of the problem's dimension. We illustrate the satisfactory performance of our proposed classifiers in both small and high dimensional settings via a simulation study and a real data example. The code implementing the proposed methods is publicly available in the {\tt R} package {\tt Qtools}.
\end{abstract}



\end{frontmatter}

\section{Introduction}
\label{sec:1}
The idea of using quantiles in classification is relatively recent and largely unexplored. The median classifier for high-dimensional problems proposed by \cite{HTX09}, which calculates the $L_{1}$ distance of the coordinates of a multivariate data point from componentwise medians (rather than centroids), is particularly advantageous when data exhibit heavy-tailed or skewed distributions. Building on \citeauthor{HTX09}'s (\citeyear{HTX09}) idea, \cite{HV2016} proposed quantile classifiers which hinge on the sum of distances from componentwise quantiles at some generic level $\theta \in (0,1)$. The ensemble quantile classifier by \cite{LM20} assigns weights to the componentwise distances by minimising a regularised loss function, where the regularisation parameter is determined by cross-validation.

In all the studies mentioned above, quantiles are calculated marginally for each input variable (componentwise). This implies that their calculation ignores the possible interdependence among variables. In this study, we consider directional quantiles for multivariate distributions \citep{KM2012} to address such a limitation. Our choice is motivated by several reasons. First, as already mentioned, the dependence among variables is taken into account by computing linear combinations of input variables. Second, directional quantiles have a simple interpretation since the projections' weights embody the relative importance of the variables involved in the classification problem. Finally, in the special case of $p$ canonical directions (with $p$ equal to the number of variables), the use of directional quantiles leads to the componentwise quantile classifier \citep{HV2016}, and thus inherits asymptotic optimal properties as shown in Appendix. Directional quantiles have already found application in risk classification problems \citep{gera:et:al:20} and proved to be a worthwhile alternative to risk classification based on componentwise quantile thresholds.

In general, the application of our methods does not require any assumption on the shape of the population distributions. We derive asymptotic theoretical properties of the proposed classifier, under the assumption that distributions for alternative populations differ by at most a location-shift. While this assumption may be unrealistic in practice, empirical results support the merit of the proposed classifier also when the distributions differ by shape and not just by location.

The rest of the paper is organised as follows. In the next section, we introduce notation and basic definitions, followed by our proposal of directional quantile classifiers. Theoretical results are stated in Section \ref{sec:3}. We report the results of a simulation study in Section \ref{sec:4} and of a real data analysis in Section \ref{sec:5}. Concluding remarks are given in Section \ref{sec:6}. All proofs of theoretical results are reported in Appendix \ref{proofs}. A software implementation of our approach can be found in the package {\tt Qtools} \citep{Qtools}, freely available on the Comprehensive R Archive Network \citep{R}.

\section{Methods}
\label{sec:2}

\subsection{Notation and definitions}
\label{sec:2.1}
Let $\mathbf{X}^{(1)}=\left(X^{(1)}_{1},X^{(1)}_{2},\ldots,X^{(1)}_{p}\right)\tp$ and $\mathbf{X}^{(2)}= \left(X^{(2)}_{1},X^{(2)}_{2},\ldots,X^{(2)}_{p}\right)\tp$ denote two $p$-variate random variables with absolutely continuous distributions $F^{(1)}$ and $F^{(2)}$ defined on the same space $\mathcal{X} \subseteq \mathbb{R}^p$ for two populations $\Pi^{(1)}$ and $\Pi^{(2)}$, respectively. The marginal distributions of the components of $\mathbf{X}^{(k)}$ are denoted by $F^{(k)}_{j}$, for $j = 1, 2, \ldots$ and $k = 1,2$. Further, $I(\cdot)$ denotes the indicator function which is equal to $1$ if its argument is true, and $0$ otherwise.

Our goal is to assign a new observation $\mathbf{y}=(y_1,y_{2},\ldots,y_p)\tp$ to either $\Pi^{(1)}$ or $\Pi^{(2)}$ according to how \textit{close} the point is to one or the other. In quantile-based classification \citep{HV2016}, the distance is first calculated for each component of $\mathbf{y}$ using the asymmetrically weighted loss function
\begin{equation}\label{e:distance}
\Phi^{(k)}(\theta; y_{j}) = \{\theta + (1-2\theta)I(y_j-Q_{X_{j}}^{(k)}(\theta) < 0)\}|y_j-Q_{X_{j}}^{(k)}(\theta)| \\
\end{equation}
for $j = 1,2,\ldots,p$ and $k = 1,2$, where $Q_{X_{j}}^{(k)}(\theta)$ is the componentwise quantile at level $\theta \in (0,1)$ for the $k$th population, which can be obtained by inversion of $F^{(k)}_{j}$. Subsequently, $\mathbf{y}$ is assigned to $\Pi^{(1)}$ if the discrepancy
\begin{equation}
\label{e:decision}
d(\mathbf{y}, \theta) = \sum_{j=1}^p \{\Phi^{(2)}(\theta; y_{j})-\Phi^{(1)}(\theta; y_{j})\}
\end{equation}
is positive, and to $\Pi^{(2)}$ otherwise. The quantile classifier reduces to the componentwise median classifier of \cite{HTX09} for $\theta=0.5$. An extension of \eqref{e:decision} to more than two populations is straightforward.

The classification rule based on~\eqref{e:decision} does not acknowledge the possible interdependence among the variables, since quantiles are obtained marginally for each variable. We address this limitation by using directional quantiles for multivariate data \citep{KM2012}. We now explain our idea informally and, in the next section, give a rigorous treatment.

Define $\mathbf{u}$ to be a vector with unit norm in $\mathbb{R}^{p}$. Throughout this paper, our focus will be on the \textit{projected} random variables $\mathbf{u}\tp \mathbf{X}^{(k)} \equiv Z^{(k)}$, $k = 1,2$, defined on $\mathcal{Z} \subseteq \mathbb{R}$. By assumption, the $Z^{(k)}$'s are continuous. We denote the corresponding distribution and density functions with $G^{(k)}(\cdot; \mathbf{u})$ and $g^{(k)}(\cdot; \mathbf{u})$, respectively.

Our goal is to develop a classifier where the quantities in~\eqref{e:distance} are opportunely redefined on the corresponding \textit{projections} along $\mathbf{u}$ to capture the multivariate nature of the distributions, namely
\begin{equation}\label{e:distance2}
\Phi^{(k)}(\theta; \mathbf{u}\tp\mathbf{y}) = \{\theta + (1-2\theta)I(\mathbf{u}\tp\mathbf{y} - Q_{Z}^{(k)}(\theta; \mathbf{u}) < 0)\}|\mathbf{u}\tp\mathbf{y} - Q_{Z}^{(k)}(\theta; \mathbf{u})|
\end{equation}
for k = 1,2, where $Q_{Z}^{(k)}(\theta; \mathbf{u}) \equiv Q_{\mathbf{u}\tp\mathbf{X}}^{(k)}(\theta)$ is the $\theta$th quantile of $Z^{(k)}$. The latter is obtained by inverting $G^{(k)}$ and it can be recognised as the $\theta$th \textit{directional quantile} of $\mathbf{X}^{(k)}$ in the direction $\mathbf{u}$ \citep{KM2012}.

\begin{figure}[!t]
\begin{center}
  \includegraphics[scale=0.8]{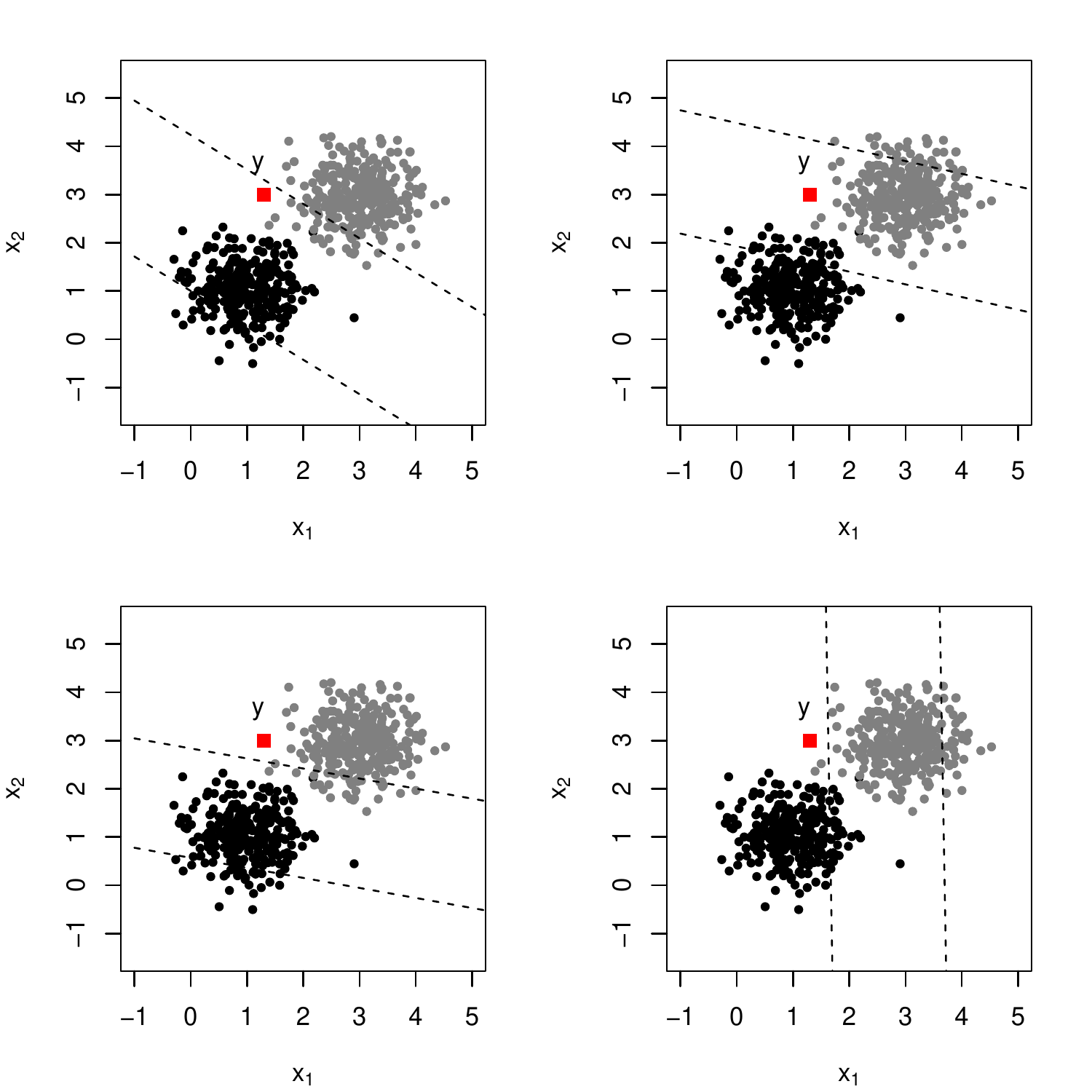}
  \caption{Simulated data depicting bivariate normal samples from two independent distributions (black and grey dots). The red filled squares mark the point with coordinates $(1.3,3.4)$, while dashed lines mark directions.}
   \label{fig:01}
  \end{center}
\end{figure}

By working with projections, we basically summarise a multivariate problem as a univariate one. Clearly, one difficulty to address is how many and which directions should be considered. To this end, we should note that not all the directions are equally useful for classification. To exemplify, consider Figure~\ref{fig:01}, which depicts bivariate normal samples from two independent populations centered at (1,1) and (3,3), respectively, and same variance. We want to assign the new observation $\mathbf{y}=(1.3,3.4)\tp$ to one of the two populations. The log-density at $\mathbf{y}$ of two bivariate normal distributions with sample means and covariance matrices separately estimated from the two samples, is $-8.8$ and $-5.7$, respectively. This suggests that $\mathbf{y}$ has been generated more likely from $F_2$ than from $F_1$.

Now compute $\Phi^{(k)}(0.9; \mathbf{u}\tp\mathbf{y})$, $k=1,2$, as in~\eqref{e:distance2} for four normalised directions. The results are reported in Table~\ref{t:table1}. Based on a principle of minimum distance, we assign $\mathbf{y}$ to $F_{2}$, thus consistently with a maximum likelihood principle, for three, though not all four, directions.

\begin{table}[!h]
\caption{Distance $\Phi^{(k)}(0.9; \mathbf{u}\tp\mathbf{y})$, $k = 1,2$, calculated for simulated data using four different directions $\mathbf{u}$.}\label{t:table1}
\centering
\begin{tabular}{rrr}
  \toprule
\multicolumn{1}{c}{$\mathbf{u}\tp$} & \multicolumn{1}{c}{$\Phi^{(1)}$} & \multicolumn{1}{c}{$\Phi^{(2)}$} \\
  \midrule
$(-0.58, -0.81) $ & 0.27 & 0.01 \\
$(0.25, 0.97) $  & 1.58 & 0.07 \\
$(-0.20, -0.98) $   & 0.30 & 0.08 \\
$(1.00, 0.02) $   & 0.03 & 0.23 \\
 \bottomrule
\end{tabular}
\end{table}

\subsection{Directional quantile classifier}
\label{sec:2.2}
Let $\vartheta = \{\theta_{1},\theta_{2},\ldots,\theta_{R}\}$ be a set of $R$ distinct quantile levels on $(0,1)$. Also, define the set $\upsilon_{r} = \{\mathbf{u}_{r1},\mathbf{u}_{r2},\ldots,\mathbf{u}_{rS_{r}}\}$ containing $S_{r}$ normalised directions associated with $\theta_{r}$, $r = 1, \ldots, R$, and let $\upsilon = \{\upsilon_{1}, \upsilon_{2}, \ldots, \upsilon_{R}\}$. (Note that for convenience one may set $S_{r} = S$ for $r = 1, \ldots, R$.)

As mentioned in the previous section, we need to be wary of particular directions that may lead us to a classification error. Therefore, we introduce weights $\omega_{rs}$ associated with each direction $\mathbf{u}_{rs}$ to decrease (or increase) their relative importance. Let $\boldsymbol\omega = (\omega_{11}, \ldots, \omega_{1S_{1}}, \ldots, \omega_{RS_{R}})\tp$ denote the vector of all such weights. We propose the discrepancy
\begin{equation}
\label{e:decision2}
d(\mathbf{y},\vartheta,\upsilon,\boldsymbol\omega) = \sum_{r=1}^{R} \sum_{s=1}^{S_{r}} \omega_{rs} \{\Phi^{(2)}(\theta_{r}; \mathbf{u}_{rs}\tp\mathbf{y})-\Phi^{(1)}(\theta_{r}; \mathbf{u}_{rs}\tp\mathbf{y})\},
\end{equation}
where $\Phi^{(k)}$ is defined in~(\ref{e:distance2}). Then our \textit{directional quantile classifier} (DQC) assigns the observation $\mathbf{y}$ to $\Pi^{(1)}$ if $d(\mathbf{y},\vartheta,\upsilon,\boldsymbol\omega)>0$, or to $\Pi^{(2)}$ otherwise. Note that if $R = 1$, $S_{r} = p$, $\omega_{rs} = 1$, and $\upsilon = \{e_{1}, e_{2}, \ldots, e_{p}\}$ the standard basis in $\mathbb{R}^{p}$, then \eqref{e:decision2} reduces to \eqref{e:decision}.

A difficulty associated with the calculation of \eqref{e:decision2} is the selection of quantile levels, directions, and weights in the training data, say $\mathbf{x}$, that give the best performance on the test data, say $\mathbf{y}$. For some prior probabilities $\pi_{1}$ and $\pi_{2}$, let
\begin{align}
\label{e:optim}
\nonumber \psi(\mathbf{x},\vartheta,\upsilon,\boldsymbol\omega)  = \, & \pi_{1} \int_{\mathcal{X}} I\{d(\mathbf{x},\vartheta,\upsilon,\boldsymbol\omega)>0\} \rds F^{(1)}(\mathbf{x})\\
& + \pi_{2} \int_{\mathcal{X}} I\{d(\mathbf{x},\vartheta,\upsilon,\boldsymbol\omega)\leq 0\} \rds F^{(2)}(\mathbf{x})
\end{align}
denote the population probability of correct classification by the DQC. Note that maximising (\ref{e:optim}) is equivalent to minimising the theoretical misclassification rate. For any given level $\theta$ and direction $\mathbf{u}$, the optimal misclassification rate is obtained when
\begin{equation*}
\pi_{1} \int_{\mathcal{X}} \Phi^{(1)}(\theta; \mathbf{u}\tp\mathbf{x}) \rds F^{(1)}(\mathbf{x}) < \pi_{1} \int_{\mathcal{X}} \Phi^{(2)}(\theta; \mathbf{u}\tp\mathbf{x}) \rds F^{(1)}(\mathbf{x})
\end{equation*}
and
\begin{equation*}
\pi_{2} \int_{\mathcal{X}} \Phi^{(2)}(\theta; \mathbf{u}\tp\mathbf{x}) \rds F^{(2)}(\mathbf{x}) < \pi_{2} \int_{\mathcal{X}} \Phi^{(1)}(\theta; \mathbf{u}\tp\mathbf{x}) \rds F^{(2)}(\mathbf{x}),
\end{equation*}
which is equivalent to minimise
\begin{align}
\label{e:teo2}
\nonumber & \pi_{1} \int_{\mathcal{X}} \left\{\Phi^{(1)}(\theta; \mathbf{u}\tp\mathbf{x})-\Phi^{(2)}(\theta; \mathbf{u}\tp\mathbf{x})\right\} \rds F^{(1)}(\mathbf{x}) \\
& + \pi_{2} \int_{\mathcal{X}} \left\{\Phi^{(2)}(\theta; \mathbf{u}\tp\mathbf{x})-\Phi^{(1)}(\theta; \mathbf{u}\tp\mathbf{x})\right\} \rds F^{(2)}(\mathbf{x}).
\end{align}

In the general problem with $K$ populations, the minimum misclassification rate is obtained when
\begin{align}\label{e:optim2}
\sum_{k=1}^{K} \pi_{k} \int_{\mathcal{X}} \Phi^{(k)}(\theta; \mathbf{u}\tp\mathbf{x}) \rds F^{(k)}(\mathbf{x}) < \sum_{k=1}^{K} \pi_{k} \int_{\mathcal{X}} \min_{k' \neq k} \Phi^{(k')}(\theta; \mathbf{u}\tp\mathbf{x}) \rds F^{(k)}(\mathbf{x}).
\end{align}
Let $\Delta^{(k)}(\mathbf{x},\theta,\mathbf{u}) = \Phi^{(k)}(\theta; \mathbf{u}\tp\mathbf{x}) -  \min_{k' \neq k} \Phi^{(k')}(\theta; \mathbf{u}\tp\mathbf{x})$. Given a sample of $n$ observations $\mathbf{x}_{i}$ and corresponding class labels $\ell_i \in \{1, 2, \ldots, K\}$, we aim to solve
\begin{equation}
\label{e:optim3}
\min_{\vartheta,\upsilon,\boldsymbol\omega} \sum_{k=1}^{K} \sum_{i: \ell_{i}=k} \sum_{r=1}^{R} \sum_{s=1}^{S_{r}} \omega_{rs}\Delta^{(k)}(\mathbf{x}_{i},\theta_{r},\mathbf{u}_{rs}).
\end{equation}
Problem \eqref{e:optim3} may seem daunting, but luckily we can solve for $\boldsymbol\omega$ rather easily. Given $\vartheta$ and $\upsilon$, problem \eqref{e:optim3} is linear with unit-norm constraints and can be minimised by using the Lagrange multiplier method. This problem has a closed-form solution given by $\hat{\boldsymbol\omega} = (\hat{\omega}_{11}, \ldots, \hat{\omega}_{1S_{1}}, \ldots, \hat{\omega}_{RS_{R}})\tp$ with generic $rs$th element
\begin{equation*}
\hat{\omega}_{rs} =\frac{\tilde{\Delta}_{rs}}{\sqrt{\sum_{r=1}^{R} \sum_{s=1}^{S_{r}} \tilde{\Delta}^{2}_{rs}}},
\end{equation*}
where $\tilde{\Delta}_{rs} = \sum_{k=1}^{K} \sum_{i : \ell_{i}=k} \Delta^{(k)}(\mathbf{x}_{i},\theta_{r},\mathbf{u}_{rs})$.

We now turn to how to choose directions and quantile levels. A crude solution would consist in doing a multidimensional grid search on $p + 1$ dimensions. However, such a solution would become computationally prohibitive even at modest values of $p$. Thankfully, we are able to mitigate the computational cost of a na\"{\i}ve numerical solution with some theoretical results (Section~\ref{sec:3}); in particular, with Theorem~\ref{theo1}, which guarantees that for each projection there exists (at least) a quantile that leads to the optimal Bayes misclassification probability, and Theorem~\ref{theo2}, which, conversely, identifies the best direction for a given quantile level. Unfortunately, a theoretical result for the simultaneous optimisation with respect to $\theta$ and $\mathbf{u}$ does not exist. Nevertheless, we show that our DQC is asymptotically optimal (i.e. the misclassification rate goes to zero) when the number of directions increases with $p$ and $n$ (Theorem~\ref{theo3}) under certain assumptions.

In summary, there are different possible approaches including randomly selecting one or more directions and using the optimal quantile levels associated with those directions; or spanning a grid of quantile levels and using the optimal directions associated with those quantiles. After some empirical investigation, we found that a strategy that gives satisfactory results in different settings is as follows. First, we define a grid of $\theta$ values spanning the unit interval and, for each of these values, randomly draw a set of normalised directions from the hyperplane that is identified as optimal according to Theorem~\ref{theo2}. The performance of a DQC based on each single $\theta$ value is evaluated using five-fold cross-validation. In the end, we use a single quantile level (optimal according to cross-validation), with the corresponding directions sampled from the optimal hyperplane. In particular, this strategy improves over the use of an asymptotically optimal quantile level when $n$ is small. Moreover, when $p$ is not too large, a similar strategy can be used to select an approximately optimal hyperplane.

\section{Theoretical results}
\label{sec:3}

In this section, we present theoretical results concerning our DQC. The proofs of lemmas and theorems are reported in the Appendix.

\subsection{Optimal quantile level $\theta$}
\label{sec:3.1}

We derive the theoretical rate of correct classification as a function of $\theta$, for given $\mathbf{u}$. We assume $K=2$ populations, although results can be generalised to $K > 2$.

\begin{lemma}\label{l:probcorrect}
 \emph{For given $\mathbf{u}$, let $Q_{\alpha}(\theta; \mathbf{u})=\min\{Q_{Z}^{(1)}(\theta; \mathbf{u}),Q_{Z}^{(2)}(\theta; \mathbf{u})\}$ with corresponding inverse $G_{\alpha}(\cdot; \mathbf{u})$, density $g_{\alpha}(\cdot; \mathbf{u})$, and prior probability of correct classification $\pi_{\alpha}$, and let $Q_{\beta}(\theta; \mathbf{u})= \max\{Q_{Z}^{(1)}(\theta; \mathbf{u}),Q_{Z}^{(2)}(\theta; \mathbf{u})\}$ with corresponding inverse $G_{\beta}(\cdot; \mathbf{u})$, density $g_{\beta}(\cdot; \mathbf{u})$, and prior probability of correct classification $\pi_{\beta}$. The probability of correct classification of the directional quantile classifier is
\begin{equation}
\label{e:theo1c}
\psi(\theta) = \pi_{\alpha} G_{\alpha}(\tilde{Q}(\theta; \mathbf{u}); \mathbf{u})+\pi_{\beta} \{1-G_{\beta}(\tilde{Q}(\theta; \mathbf{u}); \mathbf{u})\}
\end{equation}
where $\tilde{Q}(\theta; \mathbf{u})=\theta Q_{\alpha}(\theta; \mathbf{u})+(1-\theta)Q_{\beta}(\theta; \mathbf{u})$. Analogously, the theoretical misclassification rate is
\begin{equation}
\label{e:theo1d}
1-\psi(\theta) = \pi_{\alpha} \{1-G_{\alpha}(\tilde{Q}(\theta; \mathbf{u}); \mathbf{u})\}+\pi_{\beta} G_{\beta}(\tilde{Q}(\theta; \mathbf{u}); \mathbf{u}).
\end{equation}}
\end{lemma}

\begin{theorem}\label{theo1}
\emph{Assume that the density functions $g_{\alpha}(z; \mathbf{u})$ and $g_{\beta}(z; \mathbf{u})$ exist for $z$ and are nonzero on the same compact domain $\mathcal{Z}$. Further assume that there is a point $z_{0}$ with $\pi_{\alpha} g_{\alpha}(z_{0}; \mathbf{u})=\pi_{\beta} g_{\beta}(z_{0};\mathbf{u})$ so that $\pi_{\alpha} g_{\alpha}(z; \mathbf{u}) > \pi_{\beta} g_{\beta}(z; \mathbf{u})$ for $z$ on one side of $z_0$ and $\pi_{\alpha} g_{\alpha}(z; \mathbf{u}) < \pi_{\beta} g_{\beta}(z; \mathbf{u})$ for $z$ on the other side of $z_0$. Then the quantile classifier using the quantile $\tilde{Q}(\theta; \mathbf{u})$ that minimises the theoretical misclassification probability achieves the optimal Bayes misclassification probability.
}
\end{theorem}

\medskip

\begin{figure}[!t]
\begin{center}
  \includegraphics[scale=0.65]{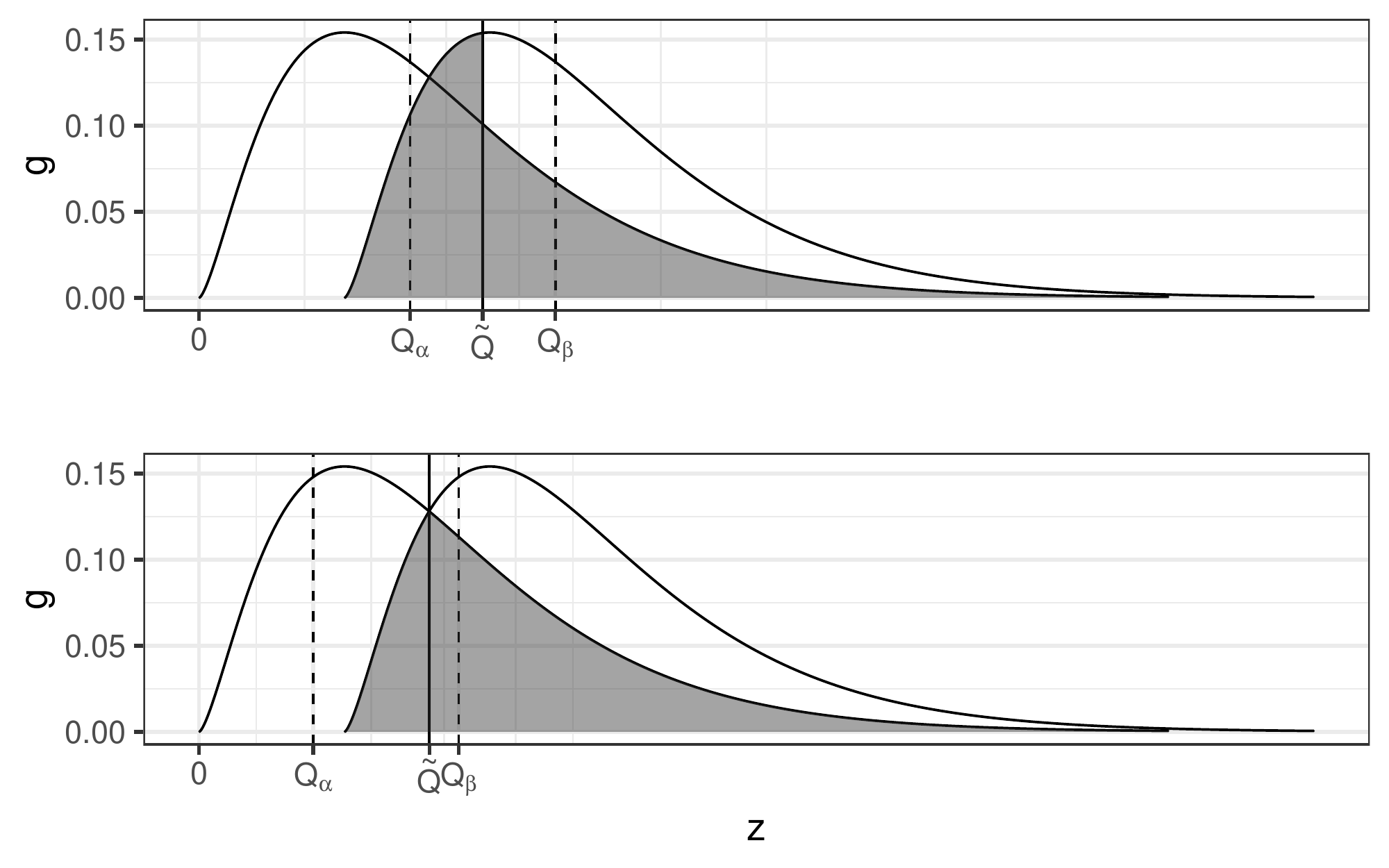}
  \caption{Misclassification probability (shaded grey area) with two location-shifted skewed distributions according to median classifier (upper panel) and the optimal quantile classifier (lower panel).}
  \label{fig:02}
  \end{center}
\end{figure}

The consistency of the classifier may be illustrated with an example. Consider a two class decision problem where one population is a location-shift version of the other. Figure \ref{fig:02} shows two distributions which have both the same right skewness. The quantiles $Q_{\alpha}(\theta)$ and $Q_{\beta}(\theta)$ are marked by dashed lines. The median classifier \citep{HTX09} in the upper panel leads to a non-optimal misclassification probability equal to 0.30. However, the misclassification probability is reduced to 0.28 by setting $\theta = 0.202$.

\subsection{Optimal direction $\mathbf{u}$}
\label{sec:3.2}
The next lemma and theorem give the optimal direction that minimises the misclassification rate at a given $\theta$.

\begin{lemma}\label{lemma2}
\emph{Let $z$ be a realisation of either $Z^{(1)}$ or $Z^{(2)}$, then
\begin{equation*}
\Phi^{(2)}(\theta; z) - \Phi^{(1)}(\theta; z) \leq Q_{Z}^{(2)}(\theta) - Q_{Z}^{(1)}(\theta),
\end{equation*}
where $\Phi^{(k)}(\theta; z) = \theta \max(\eta^{(k)},0) + (1 - \theta) \max(-\eta^{(k)},0)$ and $\eta^{(k)} = z-Q_{Z}^{(k)}(\theta)$, $k = 1,2$.
}
\end{lemma}

\begin{theorem}\label{theo2}
\emph{Let $\mathbf{W}=(W_{1},W_{2},\ldots,W_{p})\tp$ be a $p$-variate random variable such that $Q_{W_{j}}(\theta) = 0$, for $j = 1, \ldots,p$, and let $\boldsymbol\mu^{(k)}=(\mu^{(k)}_{1},\mu^{(k)}_{2},\ldots,\mu^{(k)}_{p})\tp$ be a vector of constants, $k = 1,2$. We assume that $\mathbf{X}^{(k)}= \mathbf{W} + \boldsymbol\mu^{(k)}$ and its probability distribution function is $F^{(k)}$, for $k = 1,2$. Moreover, assume that $Q_{Z}^{(2)}(\theta; \mathbf{u}) > Q_{Z}^{(1)}(\theta; \mathbf{u})$, where $Q_{Z}^{(k)}(\theta; \mathbf{u})$ is the $\theta$-quantile of $Z^{(k)} \equiv \mathbf{u}\tp \mathbf{X}^{(k)}$. (Notice that there is no loss of generality with this assumption since the case $Q_{Z}^{(2)}(\theta; \mathbf{u}) \leq Q_{Z}^{(1)}(\theta; \mathbf{u})$ can be reformulated as $Q_{Z}^{(2)}(\theta; -\mathbf{u}) > Q_{Z}^{(1)}(\theta; -\mathbf{u})$.) Under these assumptions, the normalised direction $\mathbf{u}$ that minimises the misclassification error \eqref{e:teo2} is
\begin{equation}
\label{e:maxu}
\hat{\mathbf{u}}=\frac{\boldsymbol\mu^{(2)} - \boldsymbol\mu^{(1)}}{\| \boldsymbol\mu^{(2)} - \boldsymbol\mu^{(1)}\|}
\end{equation}}
\end{theorem}

The generalization of Theorem~\ref{theo2} to $K > 2$ populations involves $K(K-1)/2$ optimal directions for each of all the possible pairwise comparisons.

\subsection{Asymptotic misclassification rate}
\label{sec:3.3}

In this section, we show that under certain assumptions, the correct classification probability converges to unity when the number of dimensions grows to infinity along with the sample size and the number of projections. The proof is built following a strategy similar to that used in \citet[][Theorem 2]{HTX09}, although our premises start from milder assumptions. In particular the projections are not required to obey the ``$\psi-mixing$ condition'' \citep{Bradley2005}, which is rather strict in practice. Our theorem is developed for any $\theta_{r} \in (0,1)$, unit weights $\omega_{rs} = 1$, and $R = 1$. Thus, the asymptotic result holds for sub-components of the summation in \eqref{e:optim3}, which are then weighted and summed to minimise the misclassification rate. Hence, the overall criterion inherits the optimal properties of its additive components.

As we did with the theorems in the previous sections, we present this theorem for $K=2$ classes. Its extension to $K > 2$ classes requires contrasting each class against the remaining $K-1$ classes, consistently with \eqref{e:optim2}.

\begin{theorem}\label{theo3}
\emph{Consider a set of directions $\upsilon = \{\mathbf{u}_{1},\ldots,\mathbf{u}_{S}\}$ sampled from a unit $p$-sphere and let $n=\max(n_{1}, n_{2})$, with $n_{1}$ and $n_{2}$ denoting the sample sizes of the two groups in the training set. Assume
\begin{enumerate}[(i)]
\item For a constant $A_{1}>0$, $S\geq A_{1} n$.
\item The $p$ variables $X_{1}^{(k)},X_{2}^{(k)},\ldots,X_{p}^{(k)}$ have each the same distribution as $W_{1}+\mu^{(k)}_{1}, W_{2}+\mu^{(k)}_{2}, \ldots, W_{p} + \mu^{(k)}_{p}$, respectively. Moreover, $Q_{W_{j}}(\theta) = 0 \ \forall j$ and $\underset{j \geq 1}{\sup} \textrm{ Var}(W_j)=A_2< + \infty$.
\item The first moments of the projections are uniformly bounded in a strong sense. This implies that $\forall c>0$ and $\forall \mathbf{u}$, $\exists \mathbf{v}$ with $|\mathbf{u}\tp\mathbf{v}|>c$ such that
    \begin{equation*}
    \underset{s \geq 1}{\inf} \, \underset{|\mathbf{u}_{s}\tp \mathbf{v}| > c}{\inf} \theta \expect |\mathbf{u}_{s}\tp \mathbf{W} + \mathbf{u}_{s}\tp \mathbf{v}| - (1-\theta) \expect |\mathbf{u}_{s}\tp \mathbf{W}| > 0.
    \end{equation*}
\item For some $\epsilon>0$, the proportion of values $s \in \{1,2,\ldots,S\}$ for which
    \begin{equation*}
    |\theta\mathbf{u}_{s}\tp \boldsymbol\mu^{(2)} - (1-\theta)\mathbf{u}_{s}\tp \boldsymbol\mu^{(1)} | > \epsilon
    \end{equation*}
    multiplied by $n^{1/2}$, say $n^{1/2}\sharp \mathcal{K}_\epsilon$, is of larger order than $S$, which means $S \left(n^{1/2} \sharp \mathcal{K}_\epsilon \right)^{-1}$ goes to zero as $n$ and $S$ increase.
\end{enumerate}
Under the previous assumptions, the directional quantile classifier $\mathcal{C}$ based on
\begin{equation*}
d(\mathbf{y},\theta,\upsilon,\boldsymbol\omega) = \sum_{s=1}^{S} \{\Phi^{(2)}(\theta; \mathbf{u}_{s}\tp\mathbf{y})-\Phi^{(1)}(\theta; \mathbf{u}_{s}\tp\mathbf{y})\},
\end{equation*}
makes the correct choice asymptotically. More specifically, as $p \rightarrow \infty$, the classifier $\mathcal{C}$ makes the correct decision with probability
\begin{equation*}
P^{(1)}\{\mathcal{C}(\mathbf{Y})=1 \} + P^{(2)}\{\mathcal{C}(\mathbf{Y})=2 \}
\end{equation*}
converging to 1 if both $n_{1}$ and $n_{2}$ diverge with $p$, where $P^{(k)}$, $k=1,2$ denotes the probability computed under the assumption that $Y$ is drawn from population $k$.
}
\end{theorem}

\section{Simulation study}
\label{sec:4}
We assessed the performance of the proposed classifier in a simulation study under three scenarios with two populations. In the first scenario, observations were generated independently from a multivariate Student's $t$ distribution with 3 degrees of freedom, with either uncorrelated or correlated variables. In the second scenario, observations were generated as in the first scenario, but each variable was subsequently transformed according to $x \mapsto  \log(|x|)$ to induce asymmetry. In both cases, the two populations differed by a location shift equal to 0.4. Finally, in the third scenario observations were generated as in the first scenario, but each variable was subsequently transformed according to $x \mapsto  \log(|x|)$ or to $x \mapsto -\log(|x|)$ depending on whether observations belonged to one or the other population, respectively.

Data were generated for each combination of overall sample size $n \in \{50, 100, 500\}$ (with $n/2$ observations in each class) and dimension $p \in \{10, 50, 100, 500\}$. All in all, this resulted in $3 \times 2 \times 3 \times 4 = 72$ simulation cases. The scale matrix used in the multivariate $t$ distribution with correlated variables was generated randomly for each $p$ using the function \texttt{rcorrmatrix} with default settings as provided in the package \texttt{clusterGeneration} \citep{clusterGeneration,Joe2006}. This resulted in non-constant pairwise correlations on the interval $(-1,1)$. Observations in the training and test datasets were generated in the same way. Data generation under each setting was replicated 100 times.

We compared the directional quantile classifier (DQC) in terms of misclassification rate on the test data with that of the centroid classifier (Centroid) \citep{THNC02}, median classifier (Median) \citep{HTX09}, componentwise quantile classifier (CQC) \citep{HV2016}, ensemble quantile classifier (EQC) \citep{LM20}, Fisher's linear discriminant analysis (LDA), $k$-nearest neighbour (KNN) \citep{CH67}, penalised logistic regression (PLR) \citep{PH08}, support vector machines (SVM) \citep{CV95,WZZ08}, and na\"{\i}ve Bayes classifier (Bayes) \citep{HY01}. Tuning parameters for PLR, KNN, and SVM where selected using cross-validation. For the CQC, the Galton correction was used to reduce skewness and optimal quantile was selected by minimising the error rate on the training set \citep{HV2016}.

We used the package \texttt{Qtools} \citep{Qtools,Qtools2} for the directional quantile classifier; the package \texttt{quantileDA} \citep{quantileDA} for the centroid, median and componentwise quantile classifiers; the package \texttt{eqc} \citep{eqc} for the ensemble quantile classifier; the package \texttt{MASS} \citep{MASS} for linear discriminant analysis; the package \texttt{class} \citep{MASS} for $k$-nearest neighbour; the package \texttt{e1071} \citep{e1071} for support vector machines and Bayes classifiers; and the package \texttt{stepPlr} \citep{stepPlr} for penalised logistic regression. All analyses were carried out in \texttt{R} version 4.0.0 \citep{R}.

The misclassification rates averaged over 100 replications for all simulation cases are reported in Tables \ref{t:sim2}-\ref{t:sim4}. The results indicate that the performance of our proposed classifier improves as $n$ and $p$ increase, in agreement with the theoretical results. In the first two scenarios, our classifier outperforms the competitors in both scenarios when variables are uncorrelated. When variables are correlated, the proposed classifier still performs very well, even if it is not uniformly the best. In the third scenario where class distributions have different shapes, the performance of our classifier is often, but not always, the best.

\begin{table}[tbp]
\centering
\small
\caption{Misclassification rates averaged over 100 replications for ten classifiers (DQC, directional quantile classifier; Centroid, centroid classifier; Median, median classifier; CQC, componentwise quantile classifier; EQC, ensemble quantile classifier; LDA, linear discriminant analysis; KNN, k-nearest neighbour; PLR, penalised logistic regression; SVM, support vector machines; Bayes, na\"{\i}ve Bayes) in the first scenario where populations have symmetric distributions.} \label{t:sim2}
\begin{tabular}{lrrrrrrrr}
\toprule
  & \multicolumn{4}{c}{\textit{Uncorrelated}} & \multicolumn{4}{c}{\textit{Correlated}}\\
\textit{Dimension} $p$ & \multicolumn{1}{c}{$10$} & \multicolumn{1}{c}{$50$} & \multicolumn{1}{c}{$100$} & \multicolumn{1}{c}{$500$} & \multicolumn{1}{c}{$10$} & \multicolumn{1}{c}{$50$} & \multicolumn{1}{c}{$100$} & \multicolumn{1}{c}{$500$} \\
\midrule
 \multicolumn{9}{c}{\textit{Sample size} $n=50$}    \\
  \midrule
DQC & 0.334 & 0.187 & 0.120 & 0.020 & 0.315 & 0.202 & 0.128 & 0.028 \\
  Centroid & 0.355 & 0.232 & 0.168 & 0.049 & 0.349 & 0.277 & 0.189 & 0.059 \\
  Median & 0.372 & 0.230 & 0.153 & 0.043 & 0.357 & 0.252 & 0.170 & 0.047 \\
  CQC & 0.362 & 0.273 & 0.220 & 0.180 & 0.367 & 0.284 & 0.222 & 0.177 \\
  EQC & 0.373 & 0.240 & 0.172 & 0.044 & 0.339 & 0.253 & 0.168 & 0.055 \\
  LDA & 0.365 & 0.382 & 0.295 & 0.313 & 0.245 & 0.252 & 0.308 & 0.339 \\
  KNN & 0.362 & 0.287 & 0.263 & 0.212 & 0.360 & 0.300 & 0.271 & 0.211 \\
  PLR & 0.348 & 0.199 & 0.134 & 0.023 & 0.275 & 0.154 & 0.103 & 0.025 \\
  SVM & 0.413 & 0.252 & 0.140 & 0.046 & 0.401 & 0.263 & 0.140 & 0.049 \\
  Bayes & 0.390 & 0.333 & 0.302 & 0.225 & 0.395 & 0.327 & 0.287 & 0.237 \\
\midrule
 \multicolumn{9}{c}{\textit{Sample size} $n=100$}    \\
  \midrule
DQC & 0.306 & 0.145 & 0.089 & 0.015 & 0.283 & 0.146 & 0.076 & 0.017 \\
  Centroid & 0.325 & 0.181 & 0.114 & 0.025 & 0.331 & 0.210 & 0.132 & 0.036 \\
  Median & 0.334 & 0.194 & 0.129 & 0.032 & 0.339 & 0.211 & 0.136 & 0.033 \\
  CQC & 0.343 & 0.213 & 0.151 & 0.076 & 0.341 & 0.223 & 0.157 & 0.092 \\
  EQC & 0.337 & 0.213 & 0.135 & 0.039 & 0.329 & 0.193 & 0.138 & 0.040 \\
  LDA & 0.338 & 0.236 & 0.393 & 0.182 & 0.226 & 0.055 & 0.105 & 0.240 \\
  KNN & 0.346 & 0.214 & 0.184 & 0.102 & 0.325 & 0.223 & 0.191 & 0.108 \\
  PLR & 0.329 & 0.182 & 0.113 & 0.019 & 0.240 & 0.092 & 0.056 & 0.020 \\
  SVM & 0.370 & 0.176 & 0.106 & 0.032 & 0.382 & 0.153 & 0.069 & 0.034 \\
  Bayes & 0.367 & 0.284 & 0.227 & 0.179 & 0.371 & 0.291 & 0.242 & 0.179 \\
\midrule
 \multicolumn{9}{c}{\textit{Sample size} $n=500$}    \\
  \midrule
DQC & 0.286 & 0.128 & 0.069 & 0.010 & 0.263 & 0.126 & 0.058 & 0.010 \\
  Centroid & 0.300 & 0.145 & 0.080 & 0.014 & 0.288 & 0.154 & 0.077 & 0.016 \\
  Median & 0.320 & 0.173 & 0.101 & 0.020 & 0.315 & 0.178 & 0.099 & 0.019 \\
  CQC & 0.327 & 0.176 & 0.108 & 0.023 & 0.320 & 0.183 & 0.103 & 0.025 \\
  EQC & 0.324 & 0.177 & 0.106 & 0.021 & 0.296 & 0.149 & 0.085 & 0.021 \\
  LDA & 0.302 & 0.160 & 0.106 & 0.367 & 0.196 & 0.027 & 0.000 & 0.036 \\
  KNN & 0.326 & 0.163 & 0.098 & 0.018 & 0.283 & 0.166 & 0.097 & 0.022 \\
  PLR & 0.301 & 0.161 & 0.104 & 0.013 & 0.194 & 0.044 & 0.018 & 0.008 \\
  SVM & 0.300 & 0.142 & 0.084 & 0.018 & 0.238 & 0.066 & 0.020 & 0.014 \\
  Bayes & 0.329 & 0.198 & 0.145 & 0.077 & 0.326 & 0.199 & 0.140 & 0.076 \\
  \toprule
\end{tabular}
\end{table}

\begin{table}[tbp]
\centering
\small
\caption{Misclassification rates averaged over 100 replications for ten classifiers (DQC, directional quantile classifier; Centroid, centroid classifier; Median, median classifier; CQC, componentwise quantile classifier; EQC, ensemble quantile classifier; LDA, linear discriminant analysis; KNN, k-nearest neighbour; PLR, penalised logistic regression; SVM, support vector machines; Bayes, na\"{\i}ve Bayes) in the second scenario where populations have distributions with same skewness.} \label{t:sim3}
\begin{tabular}{lrrrrrrrr}
\toprule
  & \multicolumn{4}{c}{\textit{Uncorrelated}} & \multicolumn{4}{c}{\textit{Correlated}}\\
\textit{Dimension} $p$ & \multicolumn{1}{c}{$10$} & \multicolumn{1}{c}{$50$} & \multicolumn{1}{c}{$100$} & \multicolumn{1}{c}{$500$} & \multicolumn{1}{c}{$10$} & \multicolumn{1}{c}{$50$} & \multicolumn{1}{c}{$100$} & \multicolumn{1}{c}{$500$} \\
\midrule
 \multicolumn{9}{c}{\textit{Sample size} $n=50$}    \\
  \midrule
DQC & 0.313 & 0.170 & 0.096 & 0.052 & 0.306 & 0.169 & 0.095 & 0.059 \\
  Centroid & 0.323 & 0.212 & 0.145 & 0.097 & 0.330 & 0.220 & 0.140 & 0.113 \\
  Median & 0.334 & 0.206 & 0.147 & 0.105 & 0.334 & 0.215 & 0.140 & 0.106 \\
  CQC & 0.350 & 0.235 & 0.187 & 0.234 & 0.360 & 0.245 & 0.193 & 0.248 \\
  EQC & 0.340 & 0.180 & 0.089 & 0.015 & 0.333 & 0.178 & 0.102 & 0.019 \\
  LDA & 0.317 & 0.383 & 0.238 & 0.228 & 0.315 & 0.397 & 0.233 & 0.237 \\
  KNN & 0.382 & 0.275 & 0.210 & 0.064 & 0.364 & 0.282 & 0.213 & 0.078 \\
  PLR & 0.313 & 0.183 & 0.095 & 0.002 & 0.322 & 0.186 & 0.098 & 0.004 \\
  SVM & 0.330 & 0.224 & 0.150 & 0.021 & 0.332 & 0.240 & 0.150 & 0.021 \\
  Bayes & 0.378 & 0.281 & 0.220 & 0.153 & 0.377 & 0.272 & 0.223 & 0.161 \\
  \midrule
 \multicolumn{9}{c}{\textit{Sample size} $n=100$}    \\
  \midrule
DQC & 0.293 & 0.129 & 0.060 & 0.012 & 0.280 & 0.118 & 0.058 & 0.010 \\
  Centroid & 0.310 & 0.168 & 0.106 & 0.057 & 0.307 & 0.161 & 0.104 & 0.065 \\
  Median & 0.328 & 0.177 & 0.110 & 0.067 & 0.316 & 0.171 & 0.106 & 0.071 \\
  CQC & 0.317 & 0.173 & 0.110 & 0.135 & 0.314 & 0.160 & 0.113 & 0.137 \\
  EQC & 0.310 & 0.135 & 0.071 & 0.006 & 0.296 & 0.126 & 0.062 & 0.006 \\
  LDA & 0.301 & 0.218 & 0.374 & 0.084 & 0.281 & 0.203 & 0.395 & 0.089 \\
  KNN & 0.358 & 0.242 & 0.188 & 0.047 & 0.353 & 0.256 & 0.186 & 0.045 \\
  PLR & 0.300 & 0.163 & 0.079 & 0.001 & 0.284 & 0.153 & 0.078 & 0.001 \\
  SVM & 0.318 & 0.177 & 0.089 & 0.005 & 0.333 & 0.168 & 0.079 & 0.005 \\
 Bayes & 0.330 & 0.229 & 0.167 & 0.095 & 0.334 & 0.225 & 0.166 & 0.098 \\
\midrule
 \multicolumn{9}{c}{\textit{Sample size} $n=500$}    \\
  \midrule
DQC & 0.273 & 0.097 & 0.038 & 0.000 & 0.265 & 0.097 & 0.035 & 0.000 \\
  Centroid & 0.282 & 0.119 & 0.059 & 0.007 & 0.275 & 0.116 & 0.056 & 0.008 \\
  Median & 0.295 & 0.128 & 0.074 & 0.017 & 0.286 & 0.124 & 0.069 & 0.019 \\
  CQC & 0.272 & 0.099 & 0.053 & 0.019 & 0.261 & 0.092 & 0.050 & 0.018 \\
  EQC & 0.267 & 0.088 & 0.035 & 0.001 & 0.244 & 0.079 & 0.032 & 0.001 \\
  LDA & 0.279 & 0.116 & 0.060 & 0.374 & 0.266 & 0.114 & 0.057 & 0.372 \\
  KNN & 0.323 & 0.206 & 0.140 & 0.016 & 0.310 & 0.207 & 0.140 & 0.015 \\
  PLR & 0.279 & 0.121 & 0.060 & 0.000 & 0.266 & 0.119 & 0.056 & 0.000 \\
  SVM & 0.283 & 0.109 & 0.046 & 0.000 & 0.274 & 0.107 & 0.044 & 0.000 \\
  Bayes & 0.273 & 0.129 & 0.080 & 0.020 & 0.266 & 0.125 & 0.077 & 0.021 \\
    \toprule
        \end{tabular}
\end{table}

\begin{table}[thbp]
\centering
\small
\caption{Misclassification rates averaged over 100 replications for ten classifiers (DQC, directional quantile classifier; Centroid, centroid classifier; Median, median classifier; CQC, componentwise quantile classifier; EQC, ensemble quantile classifier; LDA, linear discriminant analysis; KNN, k-nearest neighbour; PLR, penalised logistic regression; SVM, support vector machines; Bayes, na\"{\i}ve Bayes) in the third scenario where populations have distributions with opposite skewness.} \label{t:sim4}
\begin{tabular}{lrrrrrrrr}
\toprule
  & \multicolumn{4}{c}{\textit{Uncorrelated}} & \multicolumn{4}{c}{\textit{Correlated}}\\
\textit{Dimension} $p$ & \multicolumn{1}{c}{$10$} & \multicolumn{1}{c}{$50$} & \multicolumn{1}{c}{$100$} & \multicolumn{1}{c}{$500$} & \multicolumn{1}{c}{$10$} & \multicolumn{1}{c}{$50$} & \multicolumn{1}{c}{$100$} & \multicolumn{1}{c}{$500$} \\
\midrule
 \multicolumn{9}{c}{\textit{Sample size} $n=50$}    \\
  \midrule
DQC & 0.199 & 0.171 & 0.166 & 0.159 & 0.237 & 0.172 & 0.110 & 0.023 \\
  Centroid & 0.228 & 0.176 & 0.169 & 0.160 & 0.362 & 0.265 & 0.190 & 0.066 \\
  Median & 0.321 & 0.283 & 0.273 & 0.264 & 0.359 & 0.240 & 0.166 & 0.045 \\
  Quantile & 0.236 & 0.112 & 0.087 & 0.073 & 0.371 & 0.279 & 0.215 & 0.181 \\
  EQC & 0.315 & 0.279 & 0.256 & 0.234 & 0.349 & 0.239 & 0.162 & 0.051 \\
  LDA & 0.277 & 0.450 & 0.253 & 0.161 & 0.248 & 0.270 & 0.298 & 0.349 \\
  KNN & 0.277 & 0.213 & 0.192 & 0.173 & 0.365 & 0.284 & 0.214 & 0.074 \\
  PLR & 0.259 & 0.252 & 0.213 & 0.173 & 0.317 & 0.189 & 0.100 & 0.003 \\
  SVM & 0.231 & 0.175 & 0.170 & 0.159 & 0.338 & 0.240 & 0.157 & 0.018 \\
  Bayes & 0.229 & 0.132 & 0.123 & 0.106 & 0.373 & 0.288 & 0.227 & 0.145 \\
\midrule
 \multicolumn{9}{c}{\textit{Sample size} $n=100$}    \\
  \midrule
DQC & 0.188 & 0.162 & 0.165 & 0.166 & 0.195 & 0.133 & 0.071 & 0.016 \\
  Centroid & 0.214 & 0.167 & 0.167 & 0.166 & 0.336 & 0.212 & 0.128 & 0.033 \\
  Median & 0.314 & 0.287 & 0.285 & 0.275 & 0.341 & 0.215 & 0.132 & 0.032 \\
  Quantile & 0.214 & 0.086 & 0.071 & 0.058 & 0.346 & 0.226 & 0.159 & 0.091 \\
  EQC & 0.296 & 0.254 & 0.256 & 0.241 & 0.334 & 0.198 & 0.132 & 0.039 \\
  LDA & 0.237 & 0.300 & 0.456 & 0.174 & 0.222 & 0.056 & 0.110 & 0.238 \\
  KNN & 0.246 & 0.204 & 0.191 & 0.184 & 0.351 & 0.251 & 0.189 & 0.044 \\
  PLR & 0.234 & 0.252 & 0.235 & 0.187 & 0.284 & 0.156 & 0.079 & 0.001 \\
  SVM & 0.221 & 0.169 & 0.170 & 0.166 & 0.323 & 0.174 & 0.083 & 0.004 \\
  Bayes & 0.187 & 0.112 & 0.105 & 0.095 & 0.343 & 0.232 & 0.169 & 0.092 \\
\midrule
 \multicolumn{9}{c}{\textit{Sample size} $n=500$}    \\
  \midrule
  DQC & 0.182 & 0.166 & 0.162 & 0.159 & 0.177 & 0.111 & 0.053 & 0.010 \\
  Centroid & 0.209 & 0.170 & 0.165 & 0.160 & 0.283 & 0.153 & 0.076 & 0.015 \\
  Median & 0.312 & 0.288 & 0.283 & 0.279 & 0.312 & 0.178 & 0.099 & 0.020 \\
  Quantile & 0.203 & 0.069 & 0.052 & 0.041 & 0.316 & 0.182 & 0.104 & 0.024 \\
  EQC & 0.282 & 0.249 & 0.241 & 0.236 & 0.293 & 0.148 & 0.085 & 0.021 \\
  LDA & 0.212 & 0.194 & 0.212 & 0.474 & 0.194 & 0.027 & 0.000 & 0.032 \\
  KNN & 0.193 & 0.173 & 0.176 & 0.178 & 0.311 & 0.206 & 0.138 & 0.015 \\
  PLR & 0.213 & 0.201 & 0.226 & 0.237 & 0.266 & 0.118 & 0.057 & 0.001 \\
  SVM & 0.209 & 0.172 & 0.167 & 0.163 & 0.274 & 0.107 & 0.043 & 0.001 \\
  Bayes & 0.164 & 0.102 & 0.096 & 0.086 & 0.269 & 0.126 & 0.080 & 0.019 \\
  \toprule
\end{tabular}
\end{table}

\section{Clinical trial on Crohn's disease}
\label{sec:5}

We analyse data from a matched case-control study in first-degree relatives (FDRs) of Crohn's disease (CD) patients originally published by \cite{sorr:et:al:14}. The goal of the study was to identify asymptomatic FDRs with early CD signs using several intestinal inflammatory markers. The latter included hemoglobin, erythrocyte sedimentation rate, C-reactive protein, fecal calprotectin, and average mature ileum score. In our analysis, we grouped subjects into 2 classes, one with signs of inflammation ($n_{1} = 9$ subjects with early or frank CD) and one with normal values of markers ($n_{2} = 26$ subjects with no signs of inflammation, including healthy controls). In a separate analysis, we augment the dataset with 45 artificial markers generated from independent standard normal distributions to investigate the impact of uninformative noise on the performance of the DQC. We approach data analysis with leave-one-out validation and evaluate the misclassification rate as the proportion of subjects that are misclassified when each is left out of analysis.

We estimated the classification error for all the classifiers as included in our simulation study (Section~\ref{sec:4}). The results are reported in Table \ref{cd}.
The proposed DQC outperforms its competitors in both the original ($p = 5$) and noisy ($p = 50$) versions of the dataset.

\begin{table}[htbp]
\centering
\small
\caption{Cross-validated estimates of the misclassification rates for the Crohn's disease dataset ($p=5$) and its noisy version ($p = 50$) using ten classifiers (DQC, directional quantile classifier; Centroid, centroid classifier; Median, median classifier; CQC, componentwise quantile classifier; EQC, ensemble quantile classifier; LDA, linear discriminant analysis; KNN, $k$-nearest neighbour; PLR, penalised logistic regression; SVM, support vector machines; Bayes, na\"{\i}ve Bayes).} \label{cd}
\begin{tabular}{lrr}
\toprule
 & $p=5$ & $p=50$ \\
  \midrule
DQC & 0.229 & 0.229 \\
  Centroid & 0.286 & 0.286 \\ 
  Median & 0.400 & 0.400 \\ 
  CQC & 0.314 & 0.343 \\ 
  EQC & 0.314 & 0.314 \\ 
  LDA & 0.257 & 0.543 \\ 
  KNN & 0.371 & 0.343 \\ 
  PLR & 0.286 & 0.343 \\ 
  SVM & 0.257 & 0.257 \\ 
  Bayes & 0.286 & 0.257 \\ 
  \bottomrule
\end{tabular}
\end{table}

\section{Conclusions}
\label{sec:6}

We proposed directional quantile classifiers whose predictive ability is consistently good in both simulation and real data studies, on small and large dimensional classification problems. In particular, the empirical results show that our approach either outperforms its competitors or, when this is not the case, its performance is still in the ballpark of that of the best classifiers. Such a reliable behaviour across different scenarios is not shared by the other selected classifiers. Moreover, the directional quantile classifiers enjoy optimal theoretical properties under certain assumptions.

A limitation of the approach is that the number of directions needed to span a $p$-sphere with a regular grid becomes prohibitive already at modest values of $p$. On the other hand, our theoretical results indicate that one can sample directions from an optimal hyperplane, thus reducing the computational burden, but not at the expense of the classifier's performance. Our strategy allows us to balance the importance of quantile levels and directions used for classification by means of weights, which can be optimised using a convenient closed-form expression.

\clearpage

\appendix

\section*{Appendix A - Proofs of Theorems}
\renewcommand{\thesection}{A}
\numberwithin{equation}{section}
\setcounter{equation}{0}

\label{proofs}

\subsection{Proofs of Lemma 1 and Theorem 1}
\begin{proof}
The proofs of Lemma 1 and Theorem 1 follow the arguments given in \citet[][Supplementary Material]{HV2016}. Here, we briefly sketch the main idea. The optimal value $\theta$ that minimises the theoretical misclassification probability can be obtained by setting the first derivative of (\ref{e:theo1d}) to zero, from which
\begin{equation*}
\pi_{\alpha} g_{\alpha}\{\tilde{Q}(\theta; \mathbf{u})\} = \pi_{\beta} g_{\beta}\{\tilde{Q}(\theta; \mathbf{u})\}.
\end{equation*}

By assumption, there exists $\theta \in (0,1)$ such that $\tilde{Q}(\theta; \mathbf{u}) = z_0$. Hence, the identity above is satisfied because $Q_{\alpha}(\theta; \mathbf{u})$ and $Q_{\beta}(\theta; \mathbf{u})$ are continuous functions of $\theta$ that converge to the lower and upper bound of $\mathcal{Z}$ for $\theta$ approaching either 0 or 1, respectively. Furthermore, under the assumptions of Theorem 1, the optimal Bayesian classifier has a single decision boundary at $\tilde{Q}(\theta; \mathbf{u})$.
\end{proof}

\subsection{Proof of Lemma 2}
\begin{proof}
Without loss of generality, assume $Q_{Z}^{(1)}(\theta) \leq Q_{Z}^{(2)}(\theta)$. Let $\Delta(\theta; z) = \Phi^{(2)}(\theta; z) - \Phi^{(1)}(\theta; z)$ and consider three possible, distinct cases: $z \leq Q_{Z}^{(1)}(\theta)$, $Q_{Z}^{(1)}(\theta) < z < Q_{Z}^{(2)}(\theta)$, and $Q_{Z}^{(2)}(\theta) \leq z$.

If $z \leq Q_{Z}^{(1)}(\theta)$, then
\begin{align*}
\Delta(\theta; z) &= (1-\theta) \{Q_{Z}^{(2)}(\theta)- z\} - (1-\theta)\{Q_{Z}^{(1)}(\theta)- z\} \\
&=(1-\theta) \{Q_{Z}^{(2)}(\theta)- Q_{Z}^{(1)}(\theta)\} \leq Q_{Z}^{(2)}(\theta)- Q_{Z}^{(1)}(\theta)
\end{align*}
by definition. If $Q_{Z}^{(1)}(\theta) < z < Q_{Z}^{(2)}(\theta)$, then
\begin{align*}
\Delta(\theta; z) &= (1-\theta) \{Q_{Z}^{(2)}(\theta)-z\}-\theta \{z-Q_{Z}^{(1)}(\theta)\} \\
&= \theta\{Q_{Z}^{(1)}(\theta)- Q_{Z}^{(2)}(\theta)\} + Q_{Z}^{(2)}(\theta)-z \\
& \leq \theta Q_{Z}^{(1)}(\theta)- Q_{Z}^{(2)}(\theta) \leq Q_{Z}^{(2)}(\theta)- Q_{Z}^{(1)}(\theta).
\end{align*}
Finally, if $Q_{Z}^{(2)}(\theta) \leq z$, then
\begin{align*}
\Delta(\theta; z) &= \theta \{z-Q_{Z}^{(2)}(\theta)\}-\theta\{z-Q_{Z}^{(1)}(\theta)\}\\
&\leq Q_{Z}^{(2)}(\theta)- Q_{Z}^{(1)}(\theta).
\end{align*}
\end{proof}

\subsection{Proof of Theorem 2}

\begin{proof}
By Lemma \ref{lemma2}, the differences $\Phi^{(1)}(\theta; \mathbf{u}\tp\mathbf{x})-\Phi^{(2)}(\theta; \mathbf{u}\tp\mathbf{x})$ and $\Phi^{(2)}(\theta; \mathbf{u}\tp\mathbf{x})-\Phi^{(1)}(\theta; \mathbf{u}\tp\mathbf{x})$ are upper bounded by $Q^{(2)}_{Z}(\theta;\mathbf{u})-Q^{(1)}_{Z}(\theta;\mathbf{u})$ since $Q^{(2)}_{Z}(\theta;\mathbf{u})>Q^{(1)}_{Z}(\theta;\mathbf{u})$. Therefore the quantity in \eqref{e:teo2}, which is to be minimised with respect to $\mathbf{u}$ subject to $\|\mathbf{u}\| = 1$, is uniformly bounded above by
\begin{align*}
\label{e:theo2}
& \pi_{1} \int_{\mathcal{X}} \left\{\Phi^{(1)}(\theta; \mathbf{u}\tp\mathbf{x})-\Phi^{(2)}(\theta; \mathbf{u}\tp\mathbf{x})\right\} \rds F^{(1)}(\mathbf{x})\\
&+ \pi_{2} \int_{\mathcal{X}} \left\{\Phi^{(2)}(\theta; \mathbf{u}\tp\mathbf{x})-\Phi^{(1)}(\theta; \mathbf{u}\tp\mathbf{x})\right\} \rds F^{(2)}(\mathbf{x})  + \lambda(\mathbf{u} \tp \mathbf{u}-1)\\
&\leq  Q^{(2)}_{Z}(\theta;\mathbf{u})- Q^{(1)}_{Z}(\theta;\mathbf{u}) + \lambda(\mathbf{u}\tp \mathbf{u}-1)\\
&= (Q_{W}(\theta; \mathbf{u}) + \mathbf{u}\tp \boldsymbol\mu^{(2)}) - (Q_{W}(\theta; \mathbf{u}) + \mathbf{u}\tp \boldsymbol\mu^{(1)}) + \lambda(\mathbf{u}\tp \mathbf{u}-1)\\
&= \mathbf{u}\tp (\boldsymbol\mu^{(2)} -  \boldsymbol\mu^{(1)}) + \lambda(\mathbf{u}\tp \mathbf{u}-1).
\end{align*}
To find $\mathbf{u}$, we minimise the Lagrangian function $\mathbf{u}\tp (\boldsymbol\mu^{(2)} -  \boldsymbol\mu^{(1)}) + \lambda(\mathbf{u} \tp \mathbf{u} -1)$ which has solution given in \eqref{e:maxu}.
\end{proof}

\subsection{Proof of Theorem 3}
\begin{proof}
Let $Q_{Z}^{(k)}(\theta; \mathbf{u}_{s})$ be the empirical quantile computed on the projected training data $\mathbf{u}_{s}^\top\mathbf{X}^{(k)}$. We write
\begin{equation*}
\Phi^{(k)}(\theta; \mathbf{u}_{s}\tp\mathbf{Y}) = \gamma^{(k)}_{s}(\theta)|\mathbf{u}_{s}\tp\mathbf{Y} - Q_{Z}^{(k)}(\theta; \mathbf{u}_{s})|,
\end{equation*}
where $\gamma^{(k)}_{s}(\theta)= \theta + (1-2\theta) I\{\mathbf{u}_{s}\tp\mathbf{Y} < Q_{Z}^{(k)}(\theta; \mathbf{u}_{s})\}$. Let $\boldsymbol\mu_y$ denote the vector of quantiles of $\bf{Y}$, and put $\boldsymbol\mu_y^{(k)}=\boldsymbol\mu^{(k)}- \boldsymbol\mu_{y}$ for $k=1,2$ and write $\bf{V}=\bf{Y}-\boldsymbol\mu_{y}$. By the triangular inequality
\begin{align*}
   & \gamma^{(2)}_{s}(\theta) | \mathbf{u}_{s}\tp \mathbf{Y} - Q_{Z}^{(2)}(\theta; \mathbf{u}_{s}) | -  \gamma^{(1)}_{s}(\theta) | \mathbf{u}_{s}\tp \mathbf{Y} - Q_{Z}^{(1)}(\theta; \mathbf{u}_{s}) |  \\
     & = \gamma^{(2)}_{s}(\theta) | \mathbf{u}_{s}\tp \mathbf{V} - \mathbf{u}_{s}\tp \boldsymbol\mu_y^{(2)} | -  \gamma^{(1)}_{s}(\theta) | \mathbf{u}_{s}\tp \mathbf{V} - \mathbf{u}_{s}\tp \boldsymbol\mu_y^{(1)}| \\
     & + \tau_{2} | Q_{Z}^{(2)}(\theta; \mathbf{u}_{s}) - \mathbf{u}_{s}\tp \boldsymbol\mu^{(2)} | + \tau_{1} | Q_{Z}^{(1)}(\theta; \mathbf{u}_{s}) - \mathbf{u}_{s}\tp \boldsymbol\mu^{(1)} |,
\end{align*}
where $\tau_{1}$ and $\tau_{2}$ satisfy $|\tau_k| \leq 1$, $k = 1, 2$. Hence
\begin{align*}
T_{1} & \equiv \sum_{s=1}^{S}  \gamma^{(2)}_{s}(\theta) | \mathbf{u}_{s}\tp \mathbf{Y} - Q_{Z}^{(2)}(\theta; \mathbf{u}_{s}) | -  \gamma^{(1)}_{s}(\theta) | \mathbf{u}_{s}\tp \mathbf{Y} - Q_{Z}^{(1)}(\theta; \mathbf{u}_{s}) | \\
& = T_{2} + \tau_{1} R_1 + \tau_{2} R_2,
\end{align*}
where $T_{2}= \sum_{s=1}^{S} \gamma^{(2)}_{s}(\theta) | \mathbf{u}_{s}\tp \mathbf{V} - \mathbf{u}_{s}\tp \boldsymbol\mu_y^{(2)} | -  \gamma^{(1)}_{s}(\theta) | \mathbf{u}_{s}\tp \mathbf{V} - \mathbf{u}_{s}\tp \boldsymbol\mu_y^{(1)}|$, $R_1= \sum_{s=1}^{S} | Q_{Z}^{(1)}(\theta; \mathbf{u}_{s}) - \mathbf{u}_{s}\tp \boldsymbol\mu^{(1)} |$ and $R_2= \sum_{s=1}^{S}| Q_{Z}^{(2)}(\theta; \mathbf{u}_{s}) - \mathbf{u}_{s}\tp \boldsymbol\mu^{(2)} |$.

Given the convergence of the empirical quantiles to the respective population quantiles, it follows that
\begin{align*}
P^{(1)} (T_{1} > c_{1} - 2c_{2} S n^{-1/2} ) &\geq P^{(1)} (T_{2} > c_{1} ) - P(R_1>c_{2} S n^{-1/2} ) - P(R_2>c_{2} S n^{-1/2} ) \\
&\geq P^{(1)} (T_{2} > c_{1} ) - 2 \sum_{s=1}^S e^{-2 n_1 \delta^{(1)}_{s}}- 2 \sum_{s=1}^S e^{-2 n_2 \delta^{(2)}_{s}}
\end{align*}
for any $c_{1},c_{2}>0$, where
\begin{equation*}
\delta^{(k)}_{s}= \left[\min \left \{ F^{(k)}\left(\mathbf{u}_{s}\tp \boldsymbol\mu^{(k)} +\frac{c_2 S}{n^{1/2}} \right) - \theta, \theta - F^{(k)}\left(\mathbf{u}_{s}\tp \boldsymbol\mu^{(k)} -\frac{c_2 S}{n^{1/2}} \right) \right \} \right]^2.
\end{equation*}

Now define
\begin{equation*}
d_{s} = \expect\left\{ \gamma^{(2)}_{s}(\theta) | \mathbf{u}_{s}\tp ( \mathbf{V} - \boldsymbol\mu_y^{(2)} )| -  \gamma^{(1)}_{s}(\theta) | \mathbf{u}_{s}\tp (\mathbf{V} - \boldsymbol\mu_y^{(1)})|\right\}.
\end{equation*}

Given $\epsilon >0$, let $\mathcal{K}_\epsilon$ denote the set of indices $s \in \{1,2,\ldots,S\}$ such that
\begin{equation*}
|\gamma^{(2)}_{s}(\theta) \mathbf{u}_{s}\tp\boldsymbol\mu_2 - \gamma^{(1)}_{s}(\theta) \mathbf{u}_{s}\tp\boldsymbol\mu_1 | > \epsilon
\end{equation*}
$\forall \theta \in (0,1)$. Under the assumption that $\mathbf{Y}$ has distribution $F^{(1)}$, we have
\begin{align*}
d_{s} &= \expect \left\{\gamma^{(2)}_{s}(\theta) |\mathbf{u}_{s}\tp (\mathbf{Y} - \boldsymbol\mu_2)| -  \gamma^{(1)}_{s}(\theta) |\mathbf{u}_{s}\tp (\mathbf{Y} - \boldsymbol\mu_1)|  \right\}\\
  & = \gamma^{(2)}_{s}(\theta) \expect_{1} |\mathbf{u}_{s}\tp (\mathbf{Z} + \boldsymbol\mu_1 - \boldsymbol\mu_2)|  - \gamma^{(1)}_{s}(\theta) \expect_{1} |\mathbf{u}_{s}\tp \mathbf{Z} |,
\end{align*}
where $\expect_{1}$ is the expectation under $P^{(1)}$.

Therefore, by assumption $(iii)$ and provided $c \geq \epsilon$, we have
\begin{equation*}
\sum_{s \in \mathcal{K}_\epsilon} d_s \geq {a(c)} (\sharp \mathcal{K}_c)
\end{equation*}
where $a(c)>0$, with $a(c)=\gamma^{(2)}_{s}(\theta) \expect_{1} |\mathbf{u}_{s}\tp (\mathbf{Z} + \boldsymbol\mu_1 - \boldsymbol\mu_2)|  - \gamma^{(1)}_{s}(\theta) \expect_{1} |\mathbf{u}_{s}\tp \mathbf{Z} |$ in view of $(iii)$. As a consequence, for $\expect_{1}(T_{2})=\sum_{s=1}^{S} d_{s}$ and $\epsilon \rightarrow 0$, and $\forall c$, we have
\begin{align}\label{p:expected}
\expect_{1}(T_{2}) \geq a(c) (\sharp \mathcal{K}_c),
\end{align}
where $\sharp A$ denotes the cardinality of the set $A$.

By the Chebychev inequality and provided that $c_{1} < \frac{1}{2} \expect_{1}(T_{2})$, we have
\begin{align}\label{p:ineq}
 \nonumber P^{(1)} (T_{2} > c_{1}) &\geq 1- P^{(1)} (|T_{2} - \expect_{1}(T_{2})|> c_{1}) \geq 1 - c_{1}^{-2} \expect_{1}\{T_{2} - \expect_{1} (T_{2})\}^2\\
 &\geq  1- c_{1}^{-2}\var_{1}(T2) \geq  1- A_{2} c_{1}^{-2}S,
\end{align}
where $\var_{1}$ denotes the variance under $P^{(1)}$ and the second inequality follows from assumption $(ii)$; more specifically
\begin{align*}
 \var_{1} (T_{2}) &= \var_{1}\left\{\sum_{s=1}^S \left( \gamma^{(2)}_{s}(\theta) | \mathbf{u}_{s}\tp (\mathbf{V} - \boldsymbol\mu_y^{(2)} )| -  \gamma^{(1)}_{s}(\theta) | \mathbf{u}_{s}\tp ( \mathbf{V} - \boldsymbol\mu_y^{(1)}) |\right)\right\}\\
 &\leq \var_{1}\left\{\sum_{s=1}^S \left( \gamma^{(2)}_{s}(\theta)  \mathbf{u}_{s}\tp (\mathbf{V} - \boldsymbol\mu_y^{(2)} ) -  \gamma^{(1)}_{s}(\theta) \mathbf{u}_{s}\tp ( \mathbf{V} - \boldsymbol\mu_y^{(1)}) \right)\right\} \\
 &= \var_{1}\left\{\sum_{s=1}^S \left( \gamma^{(2)}_{s}(\theta)  \mathbf{u}_{s}\tp (\mathbf{W} + \boldsymbol\mu^{(1)}- \boldsymbol\mu^{(2)} ) -  \gamma^{(1)}_{s}(\theta) \mathbf{u}_{s}\tp  \mathbf{W}  \right)\right\} \\
 &\leq \sum_{s=1}^S A_{2} \mathbf{u}_{s}\tp \mathbf{u}_{s} + 2 \sum_{s=1}^{S-1} \sum_{s'=s+1}^S A_{2} \mathbf{u}_{s}\tp \mathbf{u}_{s'}.
\end{align*}

\cite{Stam1982Mar} proved that a uniform random variable on the sphere, $\mathbf{U} \in R^p$, converges to a standard Gaussian as $p \rightarrow \infty$. Therefore, for $S \rightarrow \infty$, by the strong law of large numbers we have
\begin{equation*}
\frac{2{\sum_{s=1}^{S-1} \sum_{s'=s+1}^{S}} A_{2} \mathbf{U}_{s}\tp \mathbf{U}_{s'}}{S(S-1)} \xrightarrow{a.s.}   A_{2} \expect(\boldsymbol\Xi_{1}\tp \boldsymbol\Xi_{2})=0,
\end{equation*}
where $\boldsymbol\Xi_{1}$ and $\boldsymbol\Xi_{2}$ are two independent standard Gaussians. This explains why the covariances become negligible in the last part of \eqref{p:ineq} as $p$ increases.

It remains to prove that $c_{1} < \frac{1}{2} \expect_{1}(T_{2})$. Consider $c_{1}=\frac{c_{3} S}{ n^{1/2}}$, where $c_{3}$ is a positive constant. By \eqref{p:expected}, the latter holds if $c_{3}S n^{-1/2}< \frac{1}{2}a(c) \mathcal{K}_c$. But this is true because it implies that
\begin{equation*}
S \left(n^{1/2} \sharp \mathcal{K}_c \right)^{-1}< \frac{1}{2}a(c) c_{3}^{-1},
\end{equation*}
where the term on the left goes to zero according to assumption $(iv)$ while $a(c) >0$, thus $c_{3}^{-1}>0$.

For $c_{1}=\frac{c_{3} S}{ n^{1/2}}$, we have
\begin{equation*}
P^{(1)} (T_{1} > c_{3} S n^{-1/2} - 2c_{2} S n^{-1/2} ) \geq 1- A_{2} \frac{n}{c_{3} ^2 S } -  2 \sum_{s=1}^S e^{-2 n_1 \delta^{(1)}_{s}}- 2 \sum_{s=1}^S e^{-2 n_2 \delta^{(2)}_{s}}.
\end{equation*}

We wish to choose $c_{3}$ and $c_{2}$ such as
\begin{align*}
P^{(1)} (T_{1} > 0 ) \geq 1- \epsilon.
\end{align*}
Therefore, we fix $\epsilon$ and choose $c_{3}$ such that $\frac{A_{2}}{c_{3}^2 A_{1}} \leq \epsilon$, where $A_{1}$ is defined in assumption $(i)$. It follows that
\begin{equation*}
\frac{A_{2} S}{c_{1}^2}=A_{2} \frac{n}{c_{3}^{2} S } \leq  \frac{A_{2}}{c_{3}^{2} A_{1} }  \leq  \epsilon.
\end{equation*}
Then we choose $c_{2}$ such that $c_{3} > 2c_{2}$ and observe that $2 \sum_{s=1}^S e^{-2 n_1 \delta^{(1)}_{s}} + 2 \sum_{s=1}^S e^{-2 n_2 \delta^{(2)}_{s}} \rightarrow 0$ for $n,S \rightarrow \infty$. Since this is true for each $\epsilon>0$, then $P^{(1)}(T_{1}>0) \rightarrow 1$, and similarly $P^{(2)}(T_{1}<0)\rightarrow 1$.
\end{proof}


\end{document}